\documentclass[aps,prb,10pt,twocolumn,groupedaddress,floatfix]{revtex4-2}

\usepackage[separate-uncertainty=true]{siunitx}
\usepackage{graphicx}
\usepackage{here}
\usepackage{amsmath}
\usepackage{nicefrac}
\usepackage{xcolor}
\usepackage{color}

\newcommand\ddfrac[2]{\frac{\displaystyle #1}{\displaystyle #2}}
\usepackage{dcolumn}
\usepackage{multirow}
\usepackage[normalem]{ulem}
\usepackage{tabto}
\DeclareSIUnit{\atpercent}{at\%}

\begin{document}
\renewcommand{\arraystretch}{2}


\title{Emergence of net magnetization by magnetic-field-biased diffusion in antiferromagnetic L1$_0$ NiMn}

\author{Nicolas Josten}
\email{Nicolas.Josten@uni-due.de}
\author{Olga Miroshkina}
\author{Sakia Noorzayee}
\author{Benjamin Zingsem}
\author{Aslı Çakır}
\author{Mehmet Acet}
\author{Ulf Wiedwald}
\author{Markus Gruner}
\author{Michael Farle}
\affiliation{Faculty of Physics and Center for Nanointegration (CENIDE), University of Duisburg-Essen, Duisburg, 47057, Germany}


\date{\today}

\begin{abstract}
NiMn is a collinear antiferromagnet with high magneto crystalline anisotropy ($K_2$=\SI{-9.7e5}{\joule\per\cubic\meter}). Through magnetic annealing of NiMn with excess Ni, strongly pinned magnetic moments emerge due to an imbalance in the distribution of Ni in the antiferromagnetic Mn-sublattices. The results are explained with a model of magnetic-field-biased diffusion, supported by \textit{ab~initio} calculations. Another observation is the oxidation of Mn at the surface, causing an enrichment of Ni in the sub-surface region. This leads to an additional ferromagnetic response appearing in the magnetization measurements, which can be removed by surface polishing.
\end{abstract}


\maketitle

\section{Introduction}
Collinear antiferromagnets are characterized by an anti-parallel alignment of neighboring magnetic moments resulting in a net magnetization of zero. Applying a magnetic field can have no effect result in canting of the moments in the field direction, or induce a spin-flop or spin-flip transition depending on the orientation and strength of the antiferromagnetic (AF) anisotropy. Net magnetic moments in antiferromagnets can also develop when there are defects in the crystal. This can be the case for surfaces \cite{PhysRevLett.79.1130}, interfaces \cite{PhysRevB.61.1315}, impurities \cite{PhysRevB.66.014430,PhysRevB.66.014431}, vacancies \cite{PhysRevB.48.1036}, and chemical disorder \cite{doi:10.1063/1.3359440}.

NiMn is a tetragonal L1$_0$ intermetallic antiferromagnet below \SI{1000}{\kelvin} and a cubic B2 paramagnet above this temperature. The N\'eel temperature is estimated to be about \SI{1070}{\kelvin} \cite{kren1968structures}, which lies above the L1$_0$ stability-range. NiMn has a strong magnetocrystalline anisotropy ($K_2$=\SI{-9.7e5}{\joule\per\cubic\meter}) \cite{sakuma1998electronic} and has been used as a pinning layer for exchange bias applications in films \cite{berkowitz1999exchange}.

When NiMn has a slight excess in Ni, uncompensated moments and ferromagnetic (FM) interactions can appear in the predominantly AF matrix and modify the magnetic properties such as observed in reference \cite{Pal}. The authors annealed a sample of NiMn with \SI{54.9}{\atpercent} Ni at different temperatures between \SI{750}{\kelvin} and \SI{790}{\kelvin} for \SI{20}{\hour} in \SI{1}{\tesla}. This led to the appearance of uncompensated moments due to biased diffusion of Ni-excess atoms within the Mn AF sublattices. They determined an activation energy of \SI{1.8}{\eV} typical for an atomic diffusion process. They also encountered the emergence of an isotropic magnetization, which they attributed to inhomogeneities containing excess Ni within the sample.

Uncompensated magnetic moments in antiferromagnets are of special interest for the understanding of exchange bias \cite{meiklejohn1956new,ohldag2003correlation}. A linear dependence between the amount of pinned moments and the strength of the exchange bias field was shown in reference \cite{guo2021evidence}. Therefore, controlling the amount of uncompensated moments would be necessary to optimize the exchange bias effect.

Here, we examine the formation of strongly pinned magnetic moments after annealing Ni$_{51.6}$Mn$_{48.4}$ in a magnetic field. For such a study, we stress the importance of inhomogeneities considered in reference \cite{Pal} and develop a method to control them and to eliminate their influence on the actual observation of pinned moments. We then extend the magnetic-field-biased diffusion-model used in \cite{Pal} to describe the origin of the strong directional pinning and perform \textit{ab~initio} calculations for supportive evidence.

\section{Equiatomic N\MakeLowercase{i}M\MakeLowercase{n} and the consequences of excess N\MakeLowercase{i}}

Around the equiatomic concentration, NiMn exhibits three crystallographic phases with changing temperature \cite{hansen1958constitution}. The high-temperature A1 phase is FCC. Below $\sim$\SI{1200}{\kelvin} it forms an ordered B2 lattice (CsCl). Then, below the martensitic transition \cite{adachi1985transformation} at $\sim$\SI{1000}{\kelvin}, the tetragonal L1$_0$ (CuAu) phase stabilizes, where NiMn is an antiferromagnet.

\begin{figure}[hbtp]
\vspace*{5mm}
\includegraphics[width=0.5\textwidth]{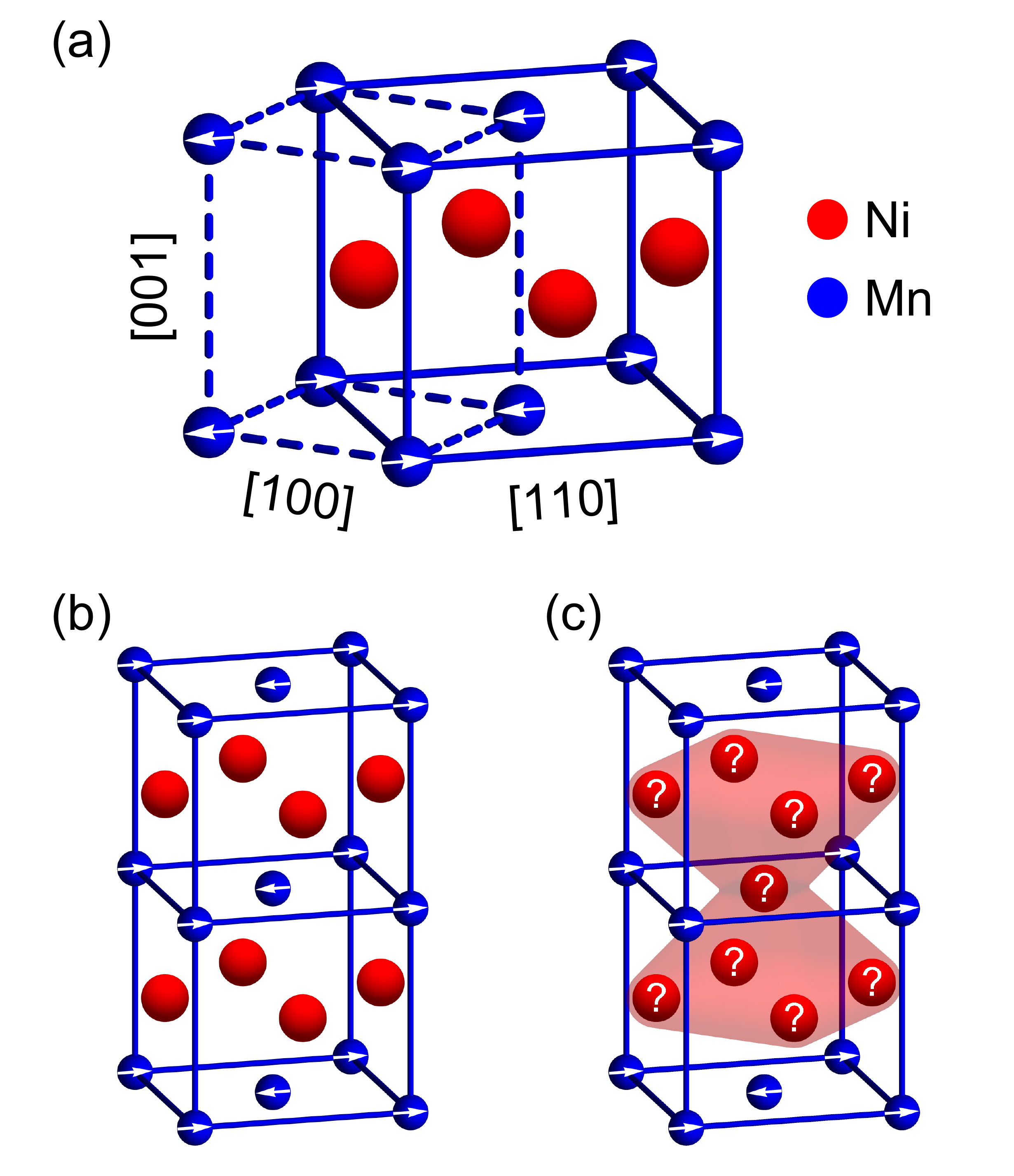}
\caption{(a) Crystal- and spin-structure of L1$_0$ NiMn. The~arrows indicate the direction of the magnetic moments. The~Ni-moment is essentially zero. (b)~The L1$_0$ NiMn crystal structure with two 'FCT' cells on top of each other. (c)~ The same structure as~(b), but with the Mn-atom at the center substituted by a Ni-atom. This leads to the formation of a nine-atom Ni-cluster shaded in red. The question marks indicate the unknown magnetic moments in this configuration.}
\label{Crystal}
\end{figure}

We show in Fig.~\ref{Crystal} the crystal- and spin-structure of L1$_0$ NiMn and the consequences of excess Ni. Figure \ref{Crystal}(a) shows as dashed lines the L1$_0$ unit cell with $P4/mmm$ space group. The structure can also be considered as a cell shown with the solid lines. This cell is described with the $C4/mmm$ space group commonly referred to as 'face-centered tetragonal'. The directions of the magnetic moments are indicated by arrows. Neutron diffraction studies show that the magnetic easy-axis lies within the basal plane. However, it cannot differentiate between the [100] and [110] directions \cite{KASPER1959231}. These studies provide a magnetic moment of 4~$\mu_{\rm B}$ per Mn atom and essentially no magnetic moment for Ni with an upper bound of 0.2~$\mu_{\rm B}$ \cite{KASPER1959231,kren1968structures}. \textit{Ab initio} calculations \cite{sakuma1998electronic} provide evidence in favor of [110] as the easy axis and a magnetic moment between 3~$\mu_{\rm B}$ and 4~$\mu_{\rm B}$ while Ni shows no moment \cite{sakuma1998electronic,spisak1999electronic}. The results of spin-polarized scanning tunneling-microscopy studies show also that the moments point along the [110] direction at the surface of NiMn films grown on Cu(001) \cite{gao2006spin}. In this work we adopt the easy-axis as the [110] direction, although it would have been just as valid if the [100] direction were adopted as the easy-axis.

Figure \ref{Crystal}(b) shows the NiMn crystal structure with two 'FCT' cells on top of each other. If excess Ni up to \SI{6}{\percent} (above which NiMn becomes FCC) is introduced, some Ni-atoms have to occupy Mn-sites. This leads to the formation of Ni-clusters with at least 9-atoms as depicted in Fig.~\ref{Crystal}(c). The question marks indicate the unknown magnetic moments in this configuration.

\section{Methods}

\subsection{Experimental}
\label{experimental}
NiMn with excess Ni was prepared by arc melting pure elements ($\geq$\SI{99.98}{\percent}) and subsequent homogenization for 5 days at \SI{1223}{\kelvin} with the sample encapsulated in a quartz tube under Ar atmosphere. Afterwards, it was quenched in water at room temperature. We used energy-dispersive X-ray spectroscopy (EDX) incorporated in a scanning electron microscope and determined the composition to be Ni$_{51.6}$Mn$_{48.4}$. Afterwards, the sample was cut into cuboids and a disk using a precision sectioning saw, while subsequent polishing removed any residues from the surface.

X-ray diffraction measurements (XRD) using Cu K$_\alpha$-radiation were carried out on the disk. The disk was then annealed again in a quartz tube with Ar for \SI{6}{\hour} at \SI{650}{\kelvin} for additional XRD measurements with and without polishing the surface.

The magnetic properties of two cuboids were measured in a vibrating sample magnetometer (VSM) using a Quantum Design PPMS DynaCool. For magnetic annealing, the samples were mounted on a Quantum Design VSM Oven Heater-Stick using ceramic-based Zircar cement. On the first sample an initial state $M(B)$-curve was measured in $\pm$ \SI{9}{\tesla} at \SI{326}{\kelvin}. Magnetic annealing was done for \SI{14.4}{\hour} at \SI{650}{\kelvin} in 5 steps in a field of \SI{9}{\tesla}, where after each step, the $M(B)$-curve was measured at \SI{326}{\kelvin}. After the last step, the surface was polished, to remove any surface corrosion, and again, the $M(B)$-curve was measured. The second cuboid was annealed for \SI{2.9}{\hour} at \SI{650}{\kelvin} without a magnetic field. Afterwards, a temperature-dependent magnetization curve at \SI{10}{\milli\tesla} was measured with a speed of \SI{4}{\kelvin\per\minute} between \SI{326}{\kelvin} and \SI{907}{\kelvin}. Then, this sample was used for XPS depth profiling to determine the chemical state and stoichiometry variations as a function of depth using a ULVAC-PHI VersaProbe II with monochromatized Al K$_\alpha$-radiation. Depth profiling was done by Ar$^+$ sputtering initially at \SI{1}{\kilo\volt} down to a depth of about \SI{25}{\nano\meter}. It was then sputtered at \SI{3}{\kilo\volt} for various times leading to a total sputtered thickness of \SI{720(30)}{\nano\meter}. The thickness was calibrated using the calibration of thermally grown SiO$_2$ on Si substrates.

\begin{figure}[h!b!t!p!]
\vspace*{5mm}
\includegraphics[width=0.45\textwidth]{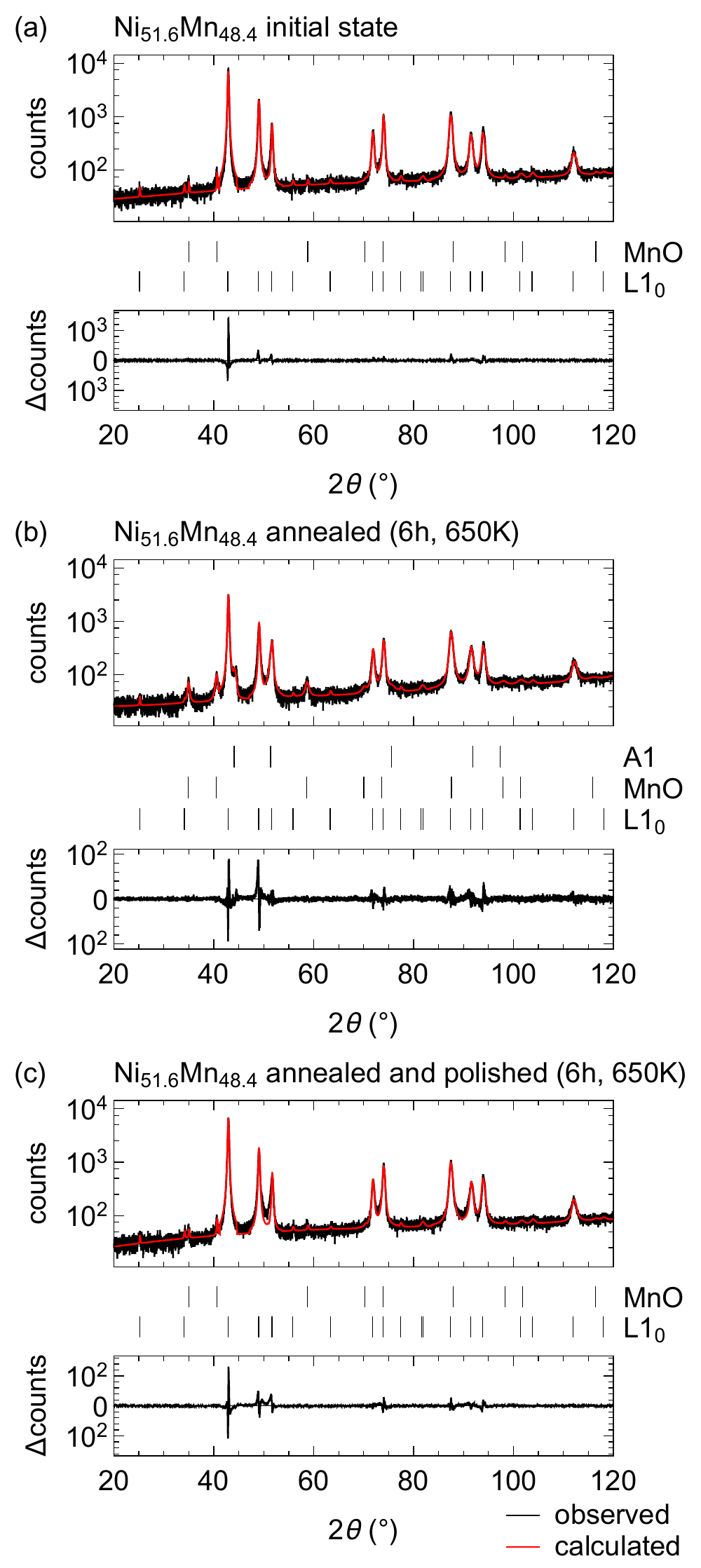}
\caption{Refined room-temperature XRD-measurements of the bulk NiMn sample. The $\Delta$counts are the difference between the observed and calculated curves. (a) XRD-measurement done directly after quenching the sample from \SI{1223}{\kelvin} in room-temperature water. (b) XRD-measurement after an annealing treatment of \SI{6}{\hour} at \SI{650}{\kelvin}. (c) XRD-measurement after additional surface polishing after the annealing treatment.}
\label{XRD}
\end{figure}

\subsection{\textit{Ab~initio} calculations}

First-principles calculations were performed with two approaches for chemical disorder modeling. The~first one is the coherent potential approximation~(CPA) realized in spin polarized relativistic Korringa-Kohn-Rostoker (SPR-KKR)~code~\cite{Ebert-code,Ebert-2011}. The~second one is a supercell approach realised in the Vienna \textit{Ab~Initio} Simulation Package~(VASP)~\cite{Kresse-1996, Kresse-1999}. For~both VASP and SPR-KKR calculations, the exchange-correlation functional was treated within the generalized gradient approximation~(GGA) following the Perdew, Burke, and Ernzerhof~(PBE)~scheme~\cite{Perdew-1996}. Detailed description of computational parameters can be found in App. \ref{Appendix}.

We~perform firstly non-collinear CPA calculations for [001], [100], and [110] spin moment directions to determine the energetically favorable configuration. Further, it also allows to evaluate the magnetocrystalline anisotropy energy~(MAE) (see App. \ref{Appendix}). In~all cases, Mn magnetic moments have a so-called layered antiparallel orientation~\cite{Kasper-1959,Siewert-2012phd,Entel-2018} which is known to be the most energetically favorable and provides zero total magnetization of~NiMn. It~was shown that the [110] spin moment direction is energetically the most favorable among the considered cases. The magnitudes of the magnetic moments are very similar for all three orientations. Results of CPA calculations are presented in App. \ref{Appendix}. For the modeling of disorder in the supercell approach, we performed collinear calculations to save computational costs. Within this approach, we modeled a 432-atom supercell by repeating a 16-atom cell three times along each of the Cartesian axes. We~considered equiatomic NiMn and a structure where one Mn atom in the middle of the supercell ((0.5; 0.5; 0.5) site) is substituted by a Ni-excess-atom. For~these systems, we performed structural relaxation and determined the total and site-resolved magnetic moments.

\section{Results}

\subsection{Structure}

We show the XRD results in Fig.~\ref{XRD}. The data were refined using JANA2006 \cite{petvrivcek2014crystallographic}. Figure~\ref{XRD}(a) shows the XRD-data of the disk-sample in the initial state. Two phases are observed. The first is the tetragonal L1$_0$ phase of NiMn, and the second is MnO. The lattice parameters we obtain for the L1$_0$ phase are $a=(2.63\pm0.01) \text{\AA}$ and $c=(3.54\pm0.01) \text{\AA}$, and for MnO $a=(4.44\pm0.01) \text{\AA}$, which are in agreement with references \cite{Pal2,ghosh2006mno}. The XRD-data after an additional heat-treatment of \SI{6}{\hour} at \SI{650}{\kelvin} is seen in Fig. \ref{XRD}(b). The amount of MnO increases, and an additional FCC phase appears. For the refinement we assume an A1 phase and obtain a lattice parameter of $a=(3.55\pm0.01) \text{\AA}$. This falls in the range of lattice parameters corresponding to stoichiometries between Ni$_{56}$Mn$_{44}$ and Ni ($a=3.52 \text{\AA}$ \cite{buschow1983magneto}). After additional polishing of the surface, no trace of the additional Ni-rich-phase is left, and the amount of MnO decreases as shown in Fig.~\ref{XRD}(c).

\subsection{Magnetization}

\label{Magnetization}
\label{surface effects}
\begin{figure*}[hbtp]
\vspace*{5mm}
\includegraphics[width=0.7\textwidth]{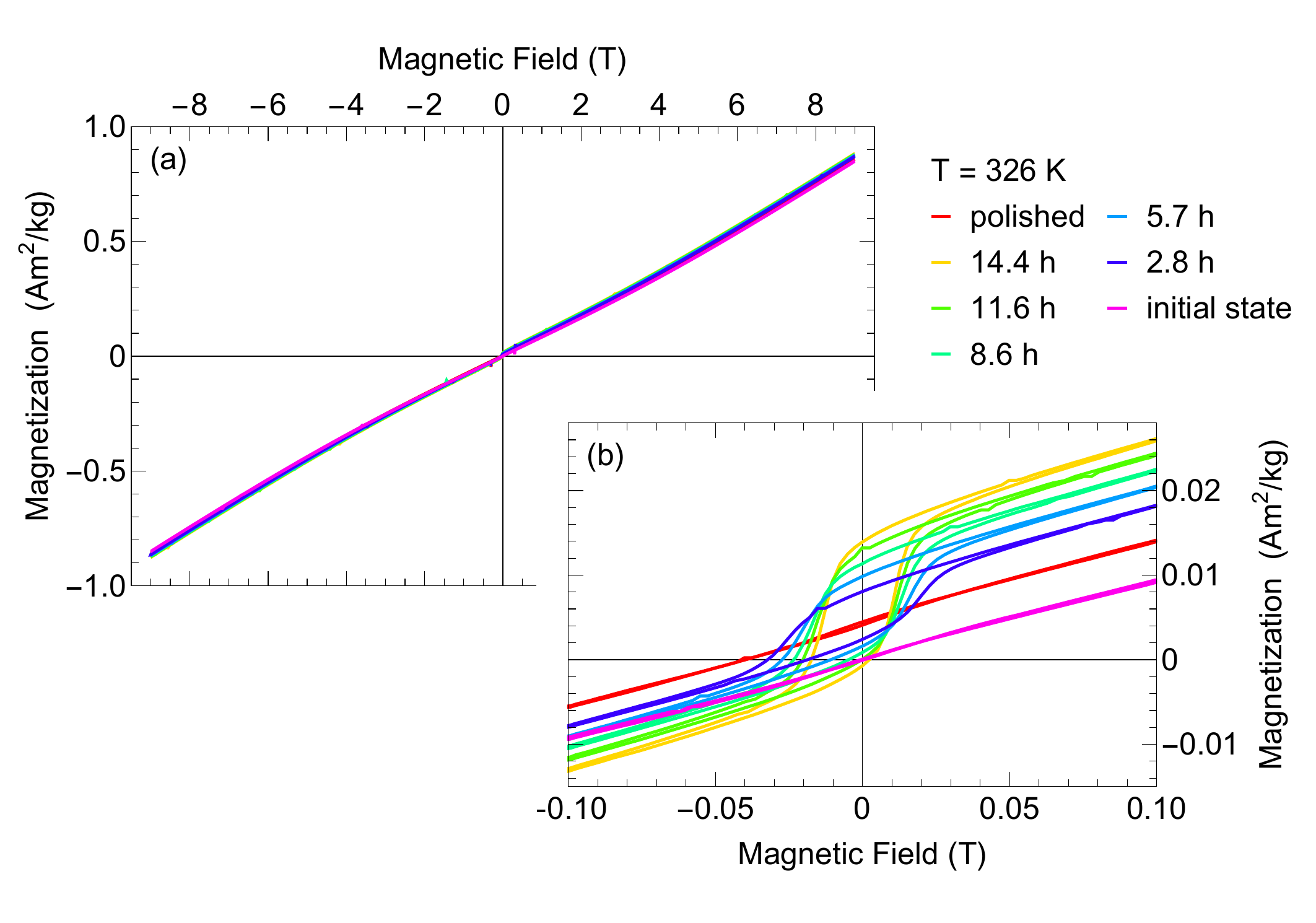}
\caption{(a) Field-dependent magnetization curves of Ni$_{51.6}$Mn$_{48.4}$ measured at \SI{326}{\kelvin} before and after magnetic annealing for various times at \SI{650}{\kelvin} and \SI{9}{\tesla}. A slight non-linearity can be observed at high fields for both measurements. (b) The region around zero field in more detail. After annealing the $M(B)$-curve is shifted upwards along the magnetization axis, which is referred to as a vertical shift. Additionally, a hysteresis emerges, which can be eliminated by polishing the sample surface.}
\label{Hyst2}
\end{figure*}

\begin{figure}[hbtp]
\vspace*{5mm}
\includegraphics[width=0.45\textwidth]{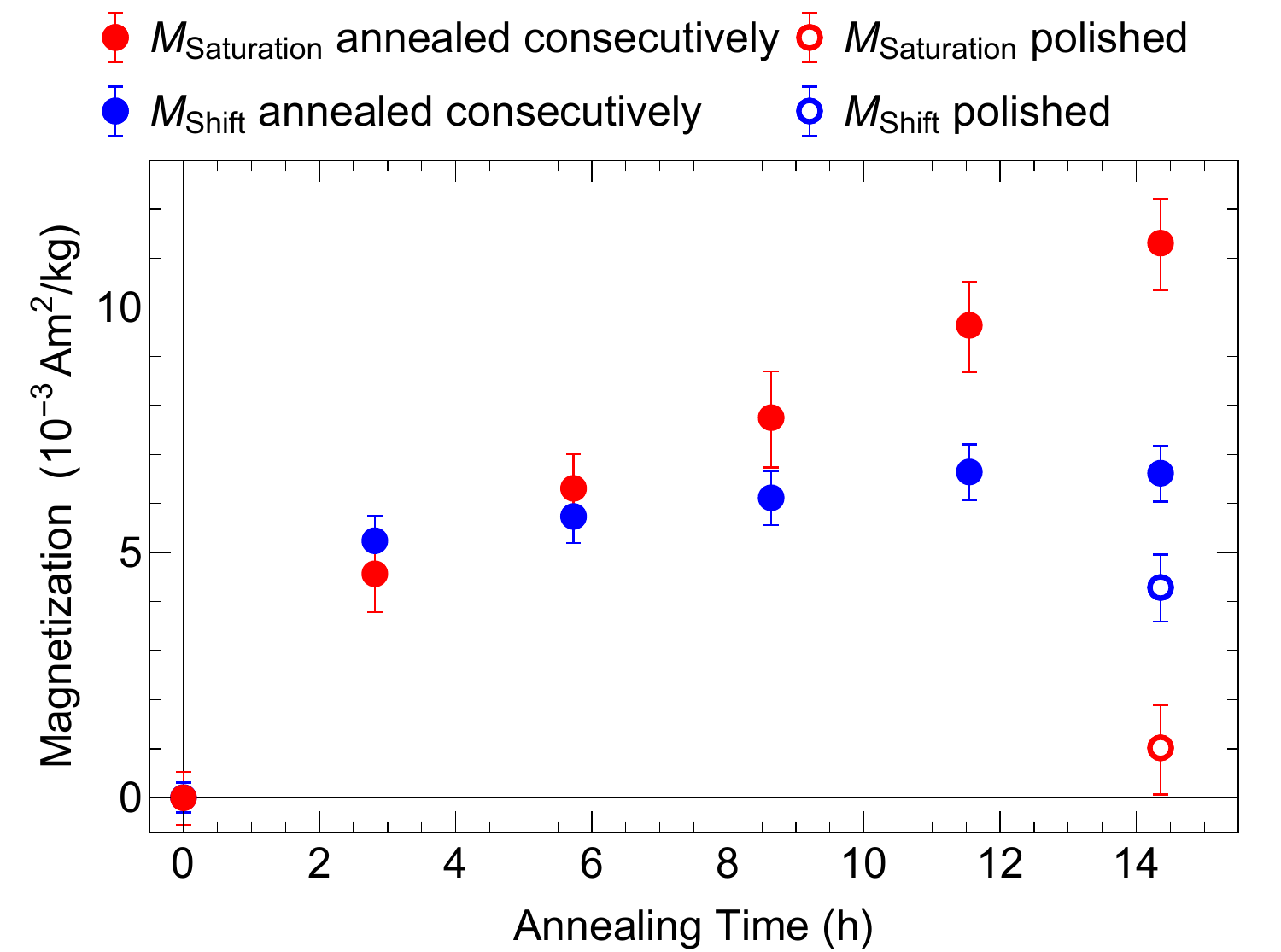}
\caption{Annealing time dependent vertical shift and saturation magnetization of the hysteresis after magnetic annealing for various times at \SI{650}{\kelvin} and \SI{9}{\tesla}. After the final annealing treatment the sample surface was polished and then remeasured. While the saturation magnetization vanishes, the vertical shift is decreased by \SI{35}{\percent}.}
\label{Shift}
\end{figure}

\begin{figure}[hbtp]
\vspace*{5mm}
\includegraphics[width=0.45\textwidth]{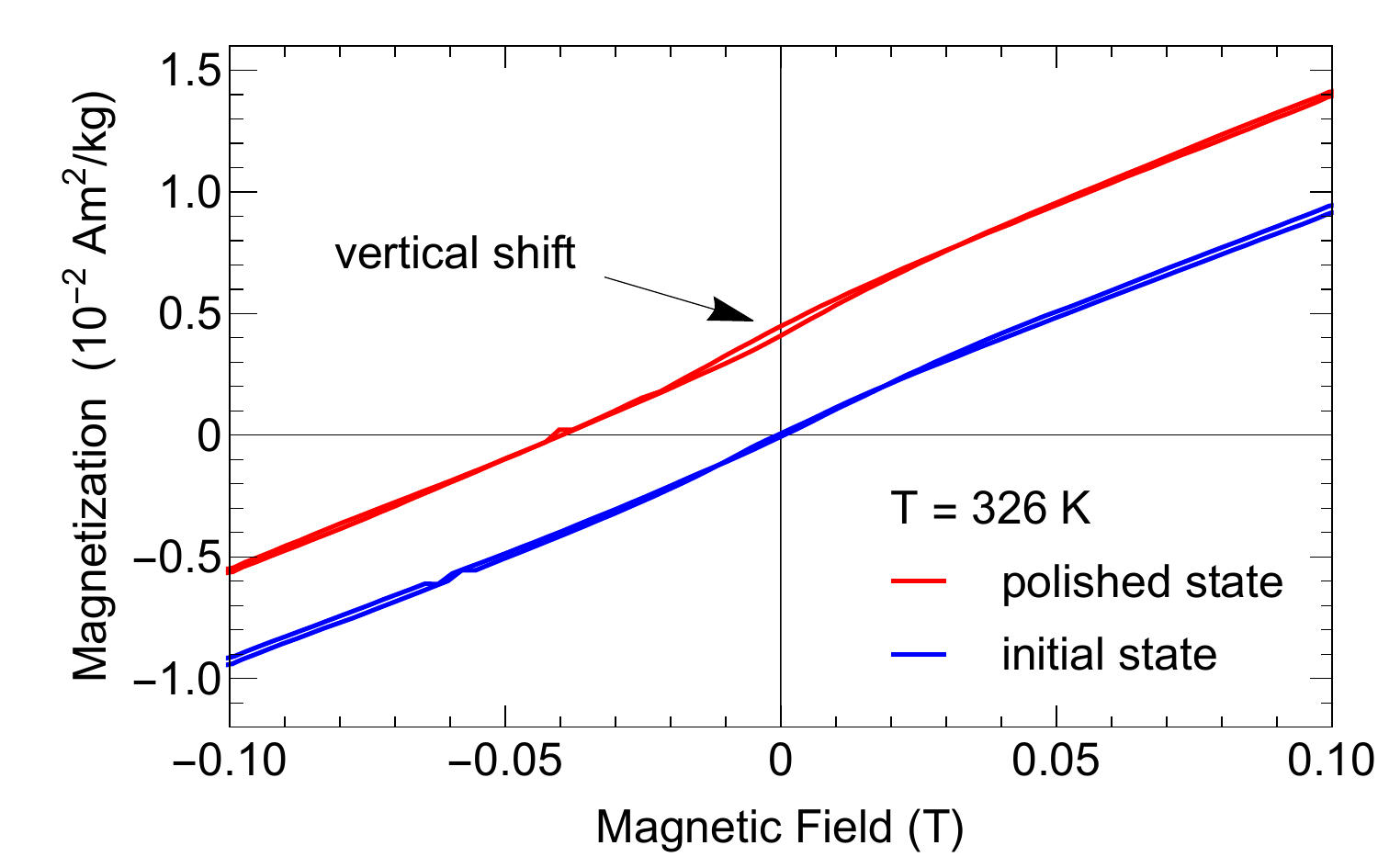}
\caption{Field-dependent magnetization measurements of Ni$_{51.6}$Mn$_{48.4}$ around zero field measured at \SI{326}{\kelvin} in the initial state and after magnetic annealing for \SI{14.4}{\hour} at \SI{650}{\kelvin} and \SI{9}{\tesla} and subsequent surface polishing.}
\label{Hyst}
\end{figure}

To be able to study the effects of magnetic-field-biased diffusion, firstly, surface and volume effects contributing to the magnetization have to be separated. Surface effects are largely eliminated by polishing the sample as mentioned in section \ref{experimental}. However, this should be verified, especially by examining in detail the surface effects themselves. We study the magnetization of initial-state (as-prepared), annealed samples, and annealed surface-treated samples to be able to account for the pinning properties occurring within the volume of the sample.

We show in Fig.~\ref{Hyst2} $M(B)$ data for Ni$_{51.6}$Mn$_{48.4}$ measured firstly in the initial state, then after consecutive annealing steps at \SI{650}{\kelvin} in a field of \SI{9}{\tesla}, and finally for the polished sample, as mentioned in section \ref{experimental}. Figure~\ref{Hyst2}(a) shows $M(B)$ in our full measurement range $\SI{-9}{\tesla}<B<\SI{9}{\tesla}$. Here, a typical behavior for an antiferromagnet can be observed with a slight non-linearity at high fields. The deviation from linearity observed as of about \SI{3}{\tesla} is related to spins being forced to rotate towards the field-direction. In this scale, the curves for all measurements overlap. However, zooming into the low-field region, as shown in Figure~\ref{Hyst2}(b), reveals features for each measurement. Two main features can be observed. Firstly, while in the initial state the $M(B)$-curve crosses the origin with no hysteresis, it develops a hysteresis essentially broadening in the vertical direction with increasing annealing time, as per measurement protocol indicated in the figure. Secondly a vertical shift of the loops also occurs even after reversing the field meaning that involved magnetic moments are strongly pinned. However, when the sample surface is polished, the hysteresis is eliminated, but the vertical shift remains.

The saturation magnetization $M_\text{Saturation}$ can be obtained by subtracting the linear response to $M(B)$ from the AF component and the vertical shift. An offset signal arising from the heater-stick of \SI{0.95e-3}{\ampere\meter^2\per\kilo\gram} was also subtracted from the saturation magnetization.

Figure \ref{Shift} shows the annealing-time dependence of the vertical shift $M_\text{Shift}$ and $M_\text{Saturation}$ along with the effect of surface polishing. Both the vertical shift and saturation magnetization increase with progressive annealing. While the vertical shift saturates gradually, the saturation magnetization increases constantly after the first annealing step. After polishing the sample surface, the contribution from the hysteresis vanishes while the vertical shift decreases by \SI{35}{\percent} as seen by the open symbols. This means that the origin of the hysteresis is related only to the surface, while the vertical shift originates from the bulk.

We now compare the $M(B)$ of the initial and polished states. For this we plot in Figure~\ref{Hyst} the respective $M(B)$ curves from fig.~\ref{Hyst2}. Here the emerging vertical shift in $M(B)$ has a value of ${M_\text{Shift}=(4.3\pm0.7)\times 10^{-3}\text{Am}^2\text{/kg}}$. All other feature in the $M(B)$-curves are identical.

\begin{figure}[hbtp]
\vspace*{5mm}
\includegraphics[width=0.45\textwidth]{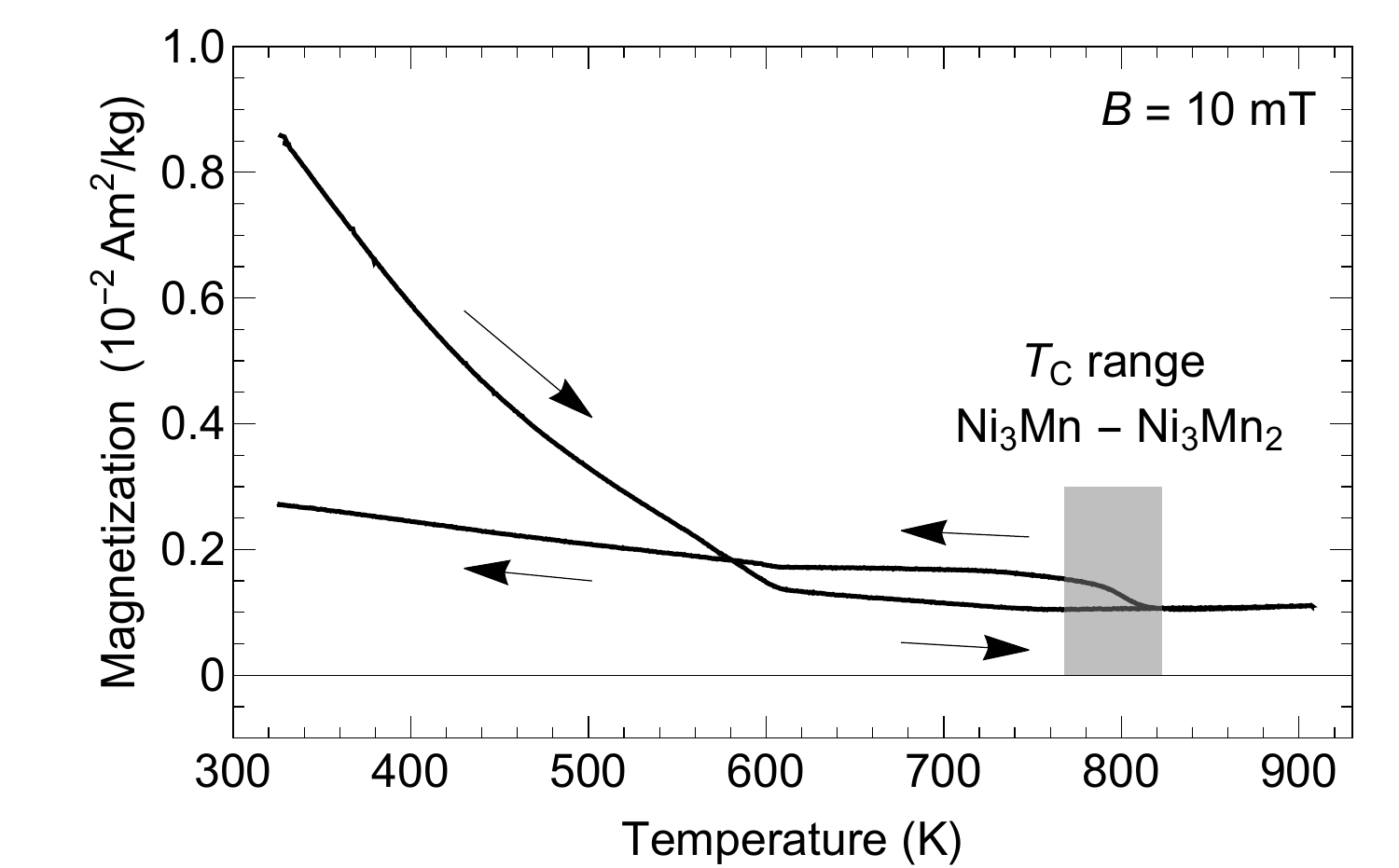}
\caption{Temperature dependent magnetization of Ni$_{51.6}$Mn$_{48.4}$ measured between \SI{326}{\kelvin} and \SI{907}{\kelvin} at \SI{10}{\milli\tesla} after annealing for \SI{2.9}{\hour} at \SI{650}{\kelvin} without a magnetic field. The magnetization declines constantly with increasing temperature until around \SI{600}{\kelvin}. While sweeping up, no distinct Curie-temperature is visible. After coming back down from \SI{907}{\kelvin} a Curie-temperature around \SI{800}{\kelvin} appears, which matches Curie-temperatures observed in stoichiometries between Ni$_3$Mn$_2$ and Ni$_3$Mn.}
\label{MTcurve}
\end{figure}

To understand the reasons behind the hysteresis occurring due to surface effects, another sample was annealed for \SI{2.9}{\hour} at \SI{650}{\kelvin} with no magnetic field applied. No vertical shift occurs in $M(B)$ in this case \cite{Pal}. This sample was not polished. We perform a temperature-dependent magnetization measurement under \SI{10}{\milli\tesla} between \SI{326}{\kelvin} and \SI{907}{\kelvin}. The small field of \SI{10}{\milli\tesla} does not lead to any substantial preferred-orientation pinning. We show the results in Fig.~\ref{MTcurve}. While sweeping the temperature upwards, a gradual decrease of magnetization is present up to \SI{600}{\kelvin}. The reason for this is the surface becomes FM when the Ni content exceeds \SI{60}{\percent} \cite{kaya1931ferromagnetismus}. The Curie-temperature $T_\text{C}$ is strongly dependent on the stoichiometry and degree of chemical order and varies between \SI{300}{\kelvin} and \SI{820}{\kelvin}. The presence of different concentrations could therefore produce a gradual decrease in $M(T)$. $M(T)$ runs flat from \SI{600}{\kelvin} up to the maximum temperature of \SI{907}{\kelvin}. When sweeping down, a sharp increase in $M(T)$ occurs around \SI{800}{\kelvin}, which corresponds to the $T_\text{C}$ range observed for compositions between ordered Ni$_{3}$Mn$_{2}$ and Ni$_{3}$Mn. This feature, which is absent in the increasing-temperature data, may be caused by homogenization and chemical ordering at the sample surface. The curve remains temperature-independent down to about \SI{600}{\kelvin}, below which it slowly increases. This corresponds to the end of the gradual decrease in $M(T)$ when measuring on increasing-temperature.

\subsection{X-ray photoemission spectroscopy}

\begin{figure}[hbtp]
\vspace*{5mm}
\includegraphics[width=0.45\textwidth]{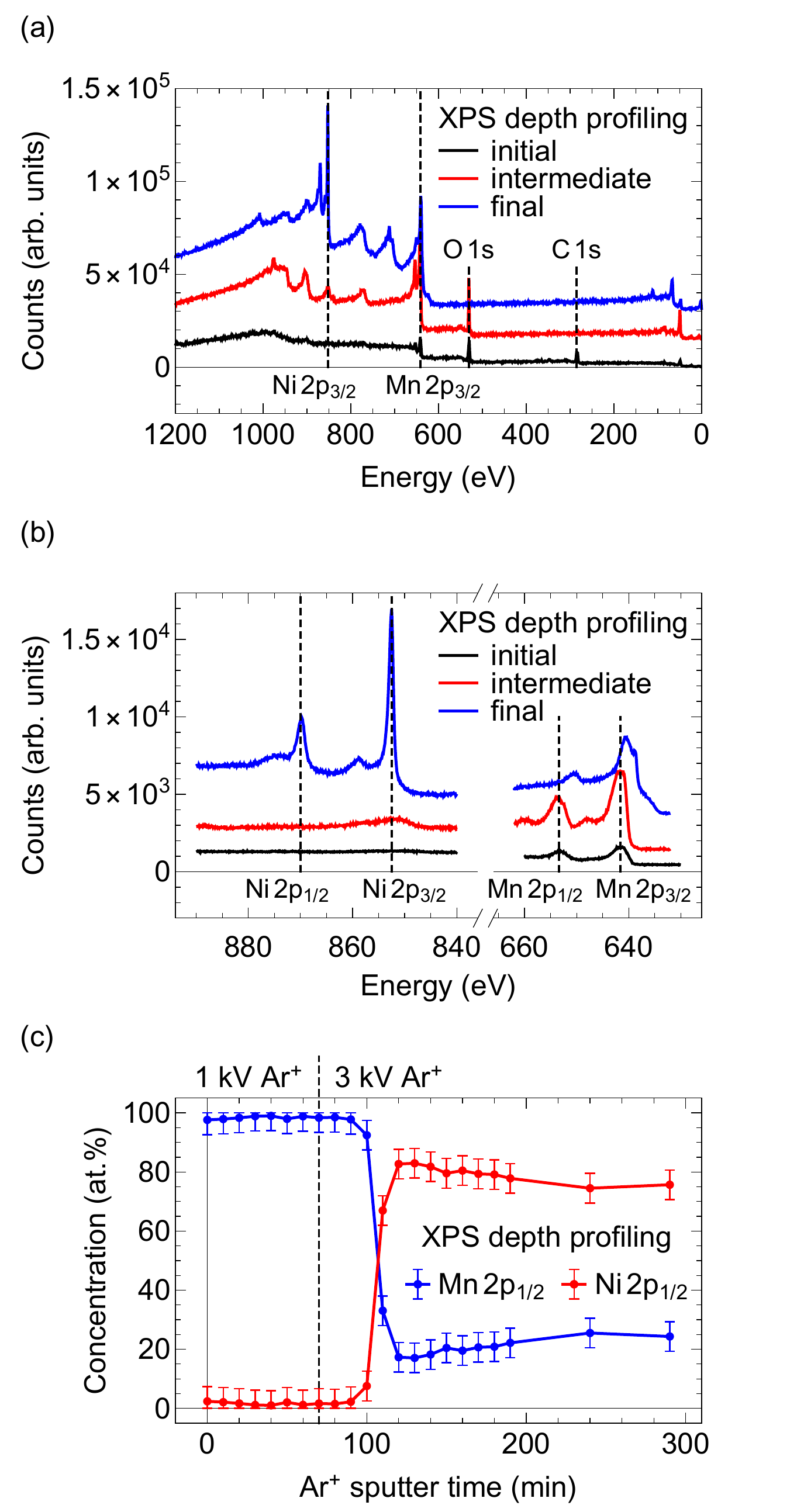}
\caption{XPS depth profiling of a bulk piece of Ni$_{51.6}$Mn$_{48.4}$ using Al-K$_\alpha$ radiation after annealing for \SI{2.9}{\hour} at \SI{650}{\kelvin} without applied field and subsequent sweep from \SI{326}{\kelvin} to \SI{907}{\kelvin} and back at a rate of \SI{4}{\kelvin\per\minute}. Survey spectra (a) and Ni-2p and Mn-2p core level spectra (b) in initial state and after sputtering with \SI{1}{\kilo\volt} Ar$^+$ ion for \SI{50}{\minute} (intermediate) and after \SI{290}{\minute} sputter time with \SI{1}{\kilo\volt} and \SI{3}{\kilo\volt} Ar$^+$ ions. (c) Relative concentrations of Ni and Mn determined from their 2p1/2 states non-overlapping with Auger peaks from the other element.}
\label{XPS}
\end{figure}

To fully understand the processes occurring at the surface, we have carried out depth resolved XPS on the sample used for the experiments plotted in Fig.~\ref{MTcurve}. Figure \ref{XPS} shows the results of XPS depth profiling.

Sputtering was carried out for \SI{70}{\minute} with \SI{1}{\kilo\volt} Ar$^+$ ions and then at \SI{3}{\kilo\volt} for \SI{220}{\minute}. Below \SI{190}{\minute} this was done in intervals of \SI{10}{\minute} and above in intervals of \SI{50}{\minute}. Full spectra up to \SI{1200}{\electronvolt} were taken in the unsputtered state (initial state), and after sputtering times of \SI{50}{\minute} (intermediate state) and \SI{290}{\minute} (final state). They are shown in Fig. \ref{XPS}(a). In the initial state, the survey spectrum shows Mn, O, and C but no Ni suggesting a significant oxidation of Mn at the surface and some C surface contaminants. The intermediate state is equivalent to about \SI{25}{\nano\meter} sputtered thickness. C is sputtered away and is thus not present in the volume. Mn and O XPS signals strengthen while that of Ni begins to emerge. In the final state, after long term sputtering of about \SI{700}{\nano\meter} sputter depth, only Mn- and Ni-related signals are present while no O is detected proving that only the metallic state is present in the volume.

The full transition from an almost Ni-free Mn oxide surface towards a metallic alloy deep in the bulk with varying stoichiometry has been tracked by the XPS 2p core level states. Figure \ref{XPS}(b) presents these spectra for the initial, intermediate and final states. For Ni-2p levels, it is clear that almost no Ni can be detected in the initial and intermediate states but a fully metallic spectrum is present in the bulk after long-time sputtering. For Mn, the intensity of 2p levels increase after sputtering without changing the energy position. Mn is fully oxidized in the intermediate state, while long-term sputtering decreases the core level energies by about \SI{3}{\electronvolt}, which corresponds to a Mn metallic signal. Note that the pronounced shoulder in the Mn-2p3/2 originates from a Ni Auger peak.

Thus, we measured the relative line intensities on the Ni- and Mn-2p1/2 peaks and determined their relative stoichiometry as shown in fig. \ref{XPS}(c). After about \SI{100}{\minute} sputtering time, the Ni-response sharply increases corresponding to an initial depth of about \SI{130}{\nano\meter}. Within the next \SI{60}{\nano\meter}, the Ni signal strongly rises up to \SI{82}{\atpercent} relative to Mn and decreases gradually to about \SI{75}{\atpercent} to the final state. In overall, the surface oxidation of the bulk Ni$_{51.6}$Mn$_{48.4}$ sample results in an almost Ni-free Mn oxide layer of about \SI{130}{\nano\meter} and a Ni-rich metallic sub-surface region. The thickness of this transition layer can be estimated to be about 2-\SI{3}{\micro\meter} by extrapolating the Ni- and Mn-content to be the volume-stoichiometry of Ni$_{51.6}$Mn$_{48.4}$. This Ni-rich metallic sub-surface region is responsible for the hysteresis contribution seen in Fig. \ref{Hyst2}. This also explains why it can be removed by surface polishing.

\begin{figure*}[hbtp]
    \centering
    \includegraphics[width=1\textwidth]{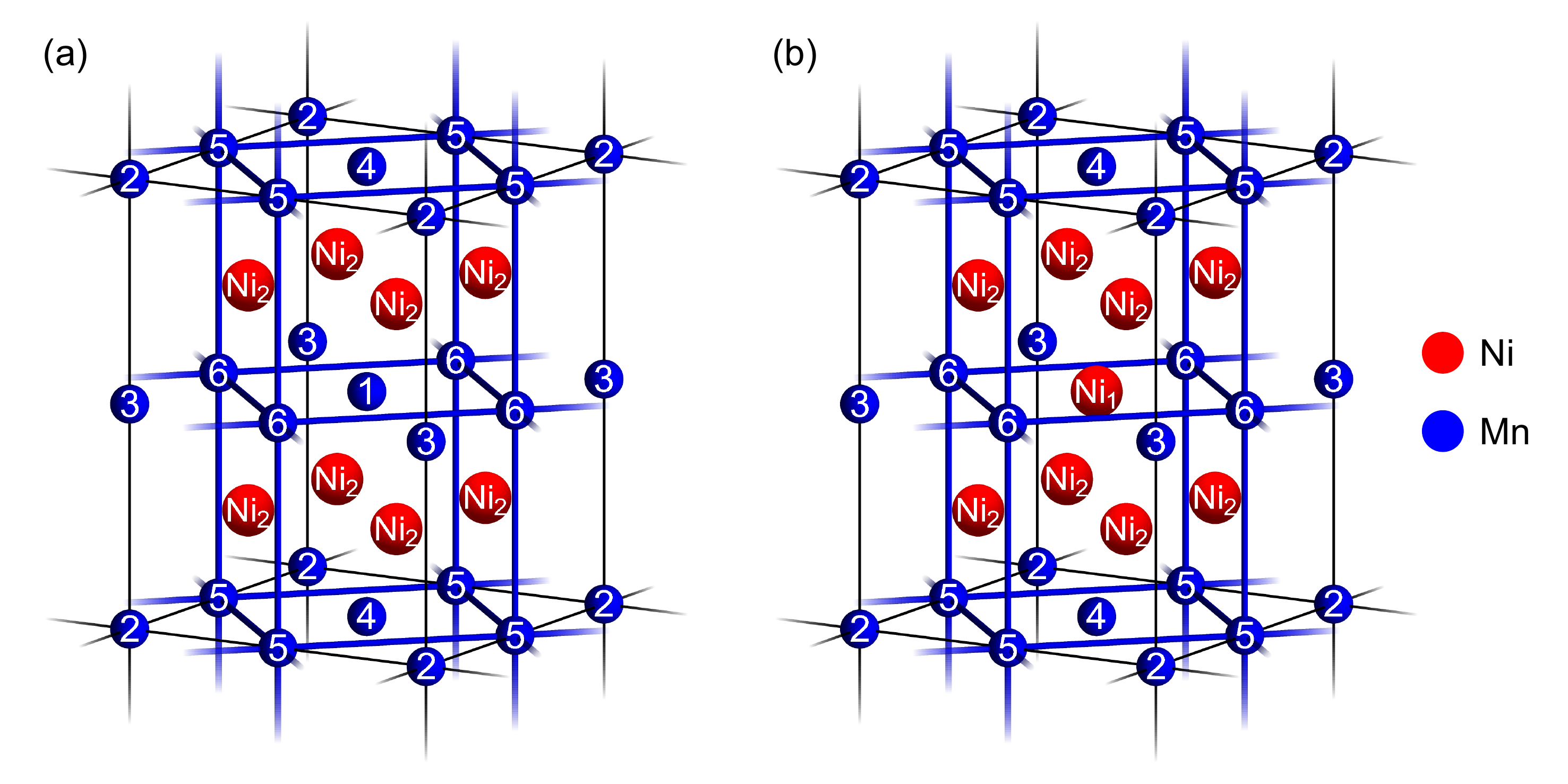}
    \caption{a)~Equiatomic NiMn and b)~structure where the~Mn-atom in the center was replaced by a Ni-atom~ (Ni$_1$). This~leads to the emergence of a non-zero net magnetic moment.
    Supercells for the first-principles calculations are formed by a $3 \times 3 \times 3$ elongation of the 16-atom cell.
    }
    \label{fig_cell}
\end{figure*}

\begin{table*}[!t]
    \caption{
   Site-resolved (in $\mu_{\rm B}$/atom) and total (in $\mu_{\rm B}$/Ni-substitution) magnetic moments~$\mu$ of the ideal equiatomic NiMn and NiMn with one Ni-excess~atom. Numbering of atoms correspond to~ Fig.~\ref{fig_cell}.
    }
    \label{table_magmom}
    \centering
    \begin{ruledtabular}
    \begin{tabular}{lccccccccc}
    \multirow{2}{*}{} & \multicolumn{8}{c}{Site-resolved $\mu$ ($\mu_\text{B}$)} &   \multirow{2}{*}{Total ($\mu_\text{B}$)} \\ \cline{2-9}
    \multicolumn{1}{c}{} & Mn$_1$ & Mn$_2$ & Mn$_3$ & Mn$_4$ & Mn$_5$ & Mn$_6$ & Ni$_1$ & Ni$_2$ &    \\ \hline
    w/o excess Ni  & -3.14 & -3.14 & -3.14 & -3.14 & 3.14 & 3.14 & -- & 0.00 & 0.00   \\
    with excess Ni   & -- & -3.13 & -3.17 & -3.16 & 3.15 & 3.22 & 0.73 & 0.09 & 5.00   \\
    \end{tabular}
    \end{ruledtabular}
\end{table*}

\subsection{Results of \textit{ab~initio} calculations}
\label{abinitio}

By \textit{ab~initio} calculations we determined how a Ni-excess atom in the Mn-plane modifies the magnetic properties. Calculated site-resolved and total magnetic moments for ideal NiMn and NiMn with one Mn-atom substituted by Ni are presented in~Table~\ref{table_magmom}.
The~cells can be found in Fig.~\ref{fig_cell}(a) and (b). When~one substitutes one Mn-atom by Ni, the magnetization changes significantly in the vicinity of this defect.
The~Ni-excess-atom (Ni$_1$ in Fig.~\ref{fig_cell}) itself acquires a moment of $\mu^{\rm{Ni_1}}=0.73$~$\mu_{\rm B}$/atom  oriented parallel to the magnetic moments of the closest Mn atoms in the plane~(Mn$_6$ in Fig.~\ref{fig_cell} with  $\mu^{\rm{Mn_6}}=3.22$~$\mu_{\rm B}$/atom).
The~eight regular Ni atoms~(Ni$_2$ in~Fig.~\ref{fig_cell}) surrounding the defect also acquire small magnetic moments of~$0.09$~$\mu_{\rm B}$/atom.
As~a result, the layered AF configuration is no longer compensated, and the structure as a whole has a net moment above zero.
In~total, one Ni~substitution generates, also including the polarization cloud involving the nearest Ni neighbors, a total magnetic moment of~5~$\mu_{\rm B}$.

\subsection{Biased diffusion}
\label{biased diffusion}

\begin{figure*}[!t]
\vspace*{5mm}
\includegraphics[width=\textwidth]{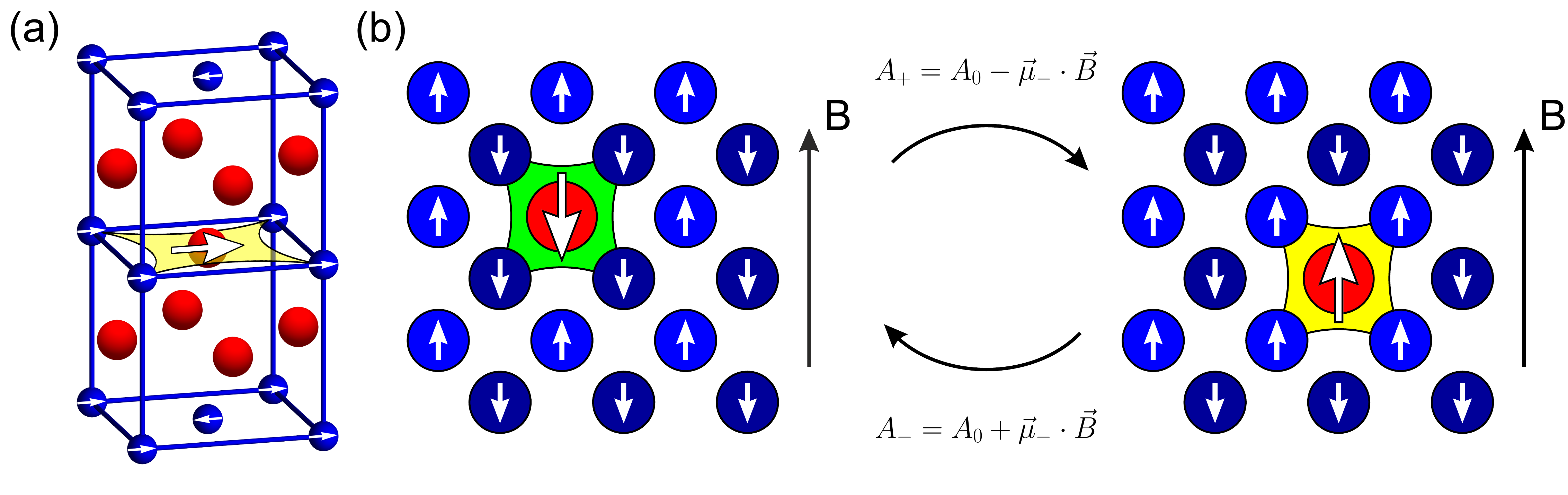}
\caption{a) L1$_0$ NiMn crystal structure from Fig.~\ref{Crystal}. The arrows again indicate the direction of the magnetic moments. The Mn-plane of interest is marked in yellow with a Ni-excess-atom in the middle. b) Schematic representation ("top view") of the biased diffusion of excess Ni in the Mn-plane due to an external applied magnetic field. The left state can be transformed into the right one and vice versa by simply moving the Ni by one atom move. $A_-$ and $A_+$ are the respective activation energies of these diffusion jumps. Here only the final states are shown and no diffusion mechanism is specified.}
\label{Diffusion}
\end{figure*}

Next, we calculate the total pinned magnetization assuming a biased diffusion of defects in single-crystalline and polycrystalline antiferromagnets after magnetic annealing. We start with
\begin{align}
    \omega = \Gamma_0 \; \text{exp} \left( - \frac{A}{k_\text{B} T} \right),
    \label{jump rate}
\end{align}
as a simple model for the diffusive jump rate $\omega$ of an atom inside a crystal lattice \cite{wert1950diffusion,balogh2014diffusion}. $\Gamma_0$ is the attempt frequency, $A$ the activation energy, $k_\text{B}$ the Boltzman constant and $T$ the temperature. In the presented case of NiMn, we concentrate on the diffusion of Ni-excess atoms within the Mn-plane. A sketch of this diffusion is given in Fig.~\ref{Diffusion}. The position of a Ni-excess-atom in the Mn-plane is highlighted yellow. In the Mn-plane the magnetic moments of neighboring Mn-atoms point in opposite directions. As demonstrated by our \textit{ab~initio} calculations (see Sec. \ref{abinitio}), the direction of the magnetic moment of an excess Ni-atom points parallel to the magnetic moments of its Mn-neighbors. A swap of position with a Mn-neighbor therefore inverts the direction of this magnetic moment. This is demonstrated in Fig.~\ref{Diffusion}(b) for a simplified case where the magnetic field and the easy axis of the antiferromagnet coincide. Therefore, the activation energy for the jump of a Ni-excess-atom can be written as
\begin{align}
    A = A_0 - \vec{\mu} \cdot \vec{B}.
    \label{activation energy}
\end{align}
Here $\vec{\mu}$ is the effective magnetic moment, which is present at the location of the Ni-excess atom occupying a Mn-site after the diffusive jump. This moment depends on the mechanism of diffusion and does not need to be the same as the moment in Tab.~\ref{table_magmom}, which we call $\vec{\mu}_\text{Total}$. This is the case, since $\vec{\mu}_\text{Total}$ represents the moment of a defect-free lattice (except for the Ni-excess atom), while different diffusion mechanisms rely on lattice defects, like vacancies. $\vec{B}$ is the applied magnetic field, and $A_0$ is the activation energy in zero field.

The Mn-plane of NiMn consists of two magnetic sublattices. This means that there only exist two possible states for Ni-excess atoms. The total number of Ni-excess atoms in the state where the Zeeman energy is minimized is $N_-$, while the number of excess atoms in the state with maximized Zeeman energy is $N_+$. The respective induced magnetic moments per excess atom are $\vec{\mu}_-$ and $\vec{\mu}_+$, so that ${\vec{\mu}_-=-\vec{\mu}_+}$. Since atoms can always switch from one state to the other, this can be treated the same way as a reversible reaction in chemistry. The reaction rates are $\omega_{-}$ and $\omega_{+}$ and their activation energies $A_{-}$ and $A_{+}$. The rates are defined as
\begin{align}
N_- \; \overset{\omega_{-}}{\underset{\omega_{+}}{\rightleftharpoons\vphantom{N}}} \; N_+.
\end{align}
If $\vec{\mu} \perp \vec{B}$, it follows that $\omega_{-}=\omega_{+}$. The jump rates can be written as
\begin{align}
\omega_{\pm} &= \Gamma_0 \; \text{exp} \left( -\frac{\left( A_0 \mp \vec{\mu}_- \cdot \vec{B}  \right) }{k_\text{B} T} \right). \label{eq omega}
\end{align}
Here the angle between $\vec{\mu}_-$ and $\vec{B}$ can only take values between $0$ and $\frac{\pi}{2}$ since $\vec{B}$ defines the direction of $\vec{\mu}_-$. The condition for equilibrium during magnetic annealing is
\begin{align}
N_- \omega_{-} &= N_+ \omega_{+}. \label{equilibrium}
\end{align}
This can be used to calculate the moment-, field- and temperature-dependent ratio $\tilde{N}$ (also called the equilibrium constant) of the occupations $N_-$ and $N_+$ as
\begin{align}
\tilde{N} &= \frac{N_-}{N_+} = \text{exp} \left( \frac{2 \;  \vec{\mu}_- \cdot \vec{B} }{k_\text{B} T} \right).
\end{align}
Both, the activation energy $A_0$ and the attempt frequency  $\Gamma_0$ cancel out on division of the rates. ${N=N_- + N_+}$ is the total number of Ni-excess atoms. ${\Delta N = N_- - N_+}$ is their difference. From the ratio $\tilde{N}$, one can calculate the relative amount of both $N_-$ and $N_+$ and the relative imbalance between $N_-$ and $N_+$ given by the equations
\begin{subequations}
\begin{eqnarray}
\frac{N_-}{N} &=&\frac{\tilde{N}}{\tilde{N}+1}, \label{relative amount 1} \\
\frac{N_+}{N} &=&\frac{1}{\tilde{N}+1}, \quad \text{and} \label{relative amount 2} \\
\frac{\Delta N}{N} &=&\frac{\tilde{N}-1}{\tilde{N}+1}. \label{relative amount 3}
\end{eqnarray}
\end{subequations}
With the atomic mass as $m_\text{Ni}$ and $m_\text{Mn}$ of Ni and Mn, respectively and the stoichometry, one can also calculate the mass magnetization $M$ of the pinned magnetization using the equations
\begin{align}
C &= \frac{1}{2} \frac{\text{Ni}_\%-\text{Mn}_\%}{\text{Ni}_\% m_\text{Ni}+\text{Mn}_\% m_\text{Mn}} \quad \text{and} \\
\begin{split}
 M &= C  \frac{\tilde{N}-1}{\tilde{N}+1}\; \vec{\mu}_\text{Total} \cdot \hat{\text{e}} \\
& = C \; \text{tanh}\left( \frac{\vec{\mu}_- \cdot \vec{B} }{k_\text{B} T} \right)\; \vec{\mu}_\text{Total} \cdot \hat{\text{e}}. \label{mass magnetization}
\end{split}
\end{align}
Here, $C$ is a scale factor to achieve the correct percentage/mass-ratio and $\hat{\text{e}}$ is a unit vector in the measurement direction. The dependence on tanh in eq. \ref{mass magnetization} is the thermal equilibrium result of a two state system. It can, for example, be found in Spin-1/2 paramagnetism \cite{aharoni2000introduction}.

Equation \ref{mass magnetization} implies that lower temperatures during annealing result in an increased pinned magnetization. While this is true for the equilibrium state, one has to consider the increased annealing-time at lower temperatures, which is not experimentally feasible.

At this point, the mentioned difference between the magnetic moment during diffusion, $\vec{\mu}_-$, and in a defect-free lattice, $\vec{\mu}_\text{Total}$, becomes important. While $\vec{\mu}_-$ has to be used to calculate the ratio $\tilde{N}$, the magnetic moment of the defect-free lattice $\vec{\mu}_\text{Total}$ is needed to calculate a mass magnetization from this ratio. Diffusion through vacancies is an example for this. It means that the Ni-atom can only swap positions with a nearest-neighbor vacancy. The vacancy itself removes a Mn-moment, which would have pointed in the same direction as the Ni-moment so that it will reduce $\mu_-$ compared to $\mu_\text{Total}$ during diffusion.

\subsection{Polycrystalline materials}
For a polycrystalline sample, the ratio of the occupation states $\tilde{N}_\text{poly}$ has to be determined by integrating the ratio of Eq. \ref{relative amount 1} and Eq. \ref{relative amount 1} and on the surface of a unit hemisphere to simulate an equal abundance of every possible crystalline direction such that
\begin{align}
\tilde{N}_\text{poly} &= \ddfrac{\frac{1}{2\pi} \int_{0}^{\frac{\pi}{2}} \int_{-\pi}^{\pi} \frac{\tilde{N}\left( \theta, \phi \right)}{\tilde{N}\left( \theta, \phi \right)+1} \text{sin} \left( \theta \right) \text{d}\phi \text{d}\theta}{\frac{1}{2\pi} \int_{0}^{\frac{\pi}{2}} \int_{-\pi}^{\pi} \frac{1}{\tilde{N}\left( \theta, \phi \right)+1} \text{sin} \left( \theta \right) \text{d}\phi \text{d}\theta}. \label{poly integral}
\end{align}
Here, $\theta$ and $\phi$ are the angles of $\vec{\mu}_-$. Without loss of generality $\vec{B}=\left(0,0,B\right)$, which simplifies the integral to
\begin{align}
\tilde{N}_\text{poly} &= \ddfrac{\int_{0}^{\frac{\pi}{2}} \ddfrac{\text{exp} \left( \gamma \;  \text{cos}\left( \theta \right) \right)}{\text{exp} \left( \gamma \;  \text{cos}\left( \theta \right) \right)+1} \text{sin} \left( \theta \right) \text{d}\theta}{ \int_{0}^{\frac{\pi}{2}} \ddfrac{1}{\text{exp} \left( \gamma \;  \text{cos}\left( \theta \right) \right)+1} \text{sin} \left( \theta \right) \text{d}\theta}. \label{simplified integral}
\end{align}
With the abbreviation
\begin{align}
\gamma = \frac{2 \mu_- B}{k_\text{B} T},
\end{align}
this integral is solved as
\begin{align}
\tilde{N}_\text{poly} &= \ddfrac{\text{ln}\left( \ddfrac{1}{2} \left(1+ \text{exp}\left(\gamma \right)\right) \right)}{\gamma + \text{ln}\left(2 \right) - \text{ln}\left(1+ \text{exp}\left(\gamma \right) \right)}, \label{poly ratio}
\end{align}
with the assumption that $\gamma \geq 0$. Now, Eq. \ref{relative amount 1} to \ref{relative amount 3} can be expressed with $\tilde{N}_\text{poly}$. To determine the mass magnetization of the pinned magnetization of a polycrystal one needs again to solve an integral on the surface of the unit hemisphere. This time, the relative excess of Ni-atoms in one of the AF sublattices is needed, and therefore, Eq. \ref{relative amount 3} has to be integrated, so that
\begin{align}
\begin{split}
     M_\text{poly} &= \frac{C \mu_\text{Total}}{2\pi} \int_{0}^{\frac{\pi}{2}} \int_{-\pi}^{\pi}
      \frac{\tilde{N}\left( \theta, \phi \right)-1}{\tilde{N}\left( \theta, \phi \right)+1} \\ & \hat{\text{e}}\left( \theta, \phi \right) \cdot \hat{\text{e}}\left( \hat{\theta}, \hat{\phi} \right)\text{sin} \left( \theta \right) \text{d}\phi \text{d}\theta.
\end{split}
\end{align}
$\hat{\theta}$ and $\hat{\phi}$ are the angles of the unit vector in themeasurement direction.
In the simplest case, $\vec{B}=\left(0,0,B\right)$ points parallel to the measurement direction, so that
\begin{align}
\begin{split}
     M_\text{poly} &= C \mu_\text{Total} \int_{0}^{\frac{\pi}{2}}
      \ddfrac{\text{exp} \left( \gamma \;  \text{cos}\left( \theta \right) \right) -1}{\text{exp} \left( \gamma \;  \text{cos}\left( \theta \right) \right)+1} \text{cos} \left( \theta \right) \; \text{sin} \left( \theta \right) \text{d}\theta \\
      &=C \mu_\text{Total} \int_{0}^{\frac{\pi}{2}}
      \text{tanh} \left( \frac{\gamma}{2} \;  \text{cos}\left( \theta \right) \right) \text{cos} \left( \theta \right) \; \text{sin} \left( \theta \right) \text{d}\theta.
\end{split}
\end{align}
From this, $M_\text{poly}$ can be calculated as
\begin{align}
\begin{split}
&M_\text{poly} = C \mu_\text{Total} \\ &\frac{\pi^2-3\gamma^2+12 \gamma \text{ln}\left(1 +\text{exp}\left(\gamma \right) \right) + 12 \text{Li}_2 \left(-\text{exp}\left(\gamma \right)  \right)}{6 \gamma^2}. \label{poly final}
\end{split}
\end{align}
Here, Li$_2$ is the polylogarithm of order 2. We note that the pinned magnetization expected from a polycrystal is one-third of the single crystal value for small values of $\gamma$, and then, with increasing $\gamma$, it approaches the expected value of half the single crystal value.

\section{Discussion}

We first compare our results with those reached in reference \cite{Pal}. Using  similar assumptions as ours, the authors determined relationships for the pinned magnetization in single-crystalline and polycrystalline NiMn. As it turns out, these relationships are the first order Taylor series expansions of Eq. \ref{mass magnetization} and Eq. \ref{poly final}, and are therefore valid for small values of $2 \mu_- B/k_\text{B} T$, which is the case for normal experimental conditions. The authors do not report a quantitative comparison between model and experiment.

Another observation the authors report is the appearance of an isotropic magnetization, which was thought to come from impurities in the sample; most likely an FCC Ni-rich Ni-Mn component. In the present study we are able to confirm this assumption and locate these impurities within the surface region. The mechanism leading to the formation of Ni-rich Ni-Mn is the oxidation of Mn at the surface leaving a Ni-rich Ni-Mn sub-surface layer. Polishing the surface removes MnO and the Ni-rich Ni-Mn and thus also the isotropic magnetization.

Now, we calculate the pinned magnetization from Eq. \ref{mass magnetization} and Eq. \ref{poly final}. For that we use the magnetic moment per Ni-excess atom $\mu_\text{Total} = 5~\mu_\text{B}$ obtained from \textit{ab~initio} calculations for NiMn (see Tab. \ref{table_magmom}). The maximum possible pinned magnetization is obtained if all Ni-excess atoms occupy only one of the two AF sublattices. For Ni$_{51.6}$Mn$_{48.4}$ this results in a value of $\SI{7.9}{\ampere\meter^2\per\kilo\gram}$. For the maximum amount of \SI{56}{\percent} of Ni in off-stoichiometric NiMn, for which the L1$_0$ structure is still stable~\cite{hansen1958constitution}, this maximum magnetization is $\SI{29.4}{\ampere\meter^2\per\kilo\gram}$. To reach these values, one has to anneal the sample infinitely long at low temperatures as implied by Eq. \ref{mass magnetization}.

Now, we provide a quantitative comparison between the model we use and the experimental results shown in Fig. \ref{Hyst}. In the experiment, the sample was annealed for \SI{14.4}{\hour} at \SI{650}{\kelvin} in \SI{9}{\tesla}, for which $M_\text{Shift}$ has saturated and reached equilibrium (Fig. \ref{Shift}). We use Eq. \ref{poly final} to calculate  the pinned magnetization of a polycrystal (basically $M_\text{Shift}$). First, we assume that $\mu_- = \mu_\text{Total}$. This gives a total calculated magnetization of $\SI{0.12}{\ampere\meter^2\per\kilo\gram}$, which is two orders of magnitude larger than the measured value of ${(4.3\pm0.7)\times 10^{-3}\text{Am}^2\text{/kg}}$.

One reason for this overestimation is possibly due to different diffusion mechanisms, which lead to a change in $\mu_-$. If diffusion is mediated through mono-vacancies, then the vacancy itself replaces another Mn-moment from the lattice and reduces $\mu_-$ compared to $\mu_\text{Total}$. $\mu_-$ is then only the moment of the additional Ni-atom, which is calculated to be $0.73~\mu_\text{B}$. If $\mu_\text{Total}$ is then again assumed to be $5~\mu_\text{B}$, Eq. \ref{poly final} results in a total magnetization of \SI{1.78e-2}{\ampere\meter^2\per\kilo\gram}, which is still four times larger than the measured value of ${(4.3\pm0.7)\times 10^{-3}\text{Am}^2\text{/kg}}$. To reach this value, $\mu_-$ would have to be $0.18~\mu_\text{B}$.

\section{Conclusions}

Annealing NiMn with excess Ni in an external field leads to a vertical shift in the magnetic field dependent magnetization curve. This shift originates from an imbalance of excess Ni atoms on Mn sites. Due to the magnetic field applied during annealing, one of the antiferromagnetic sublattices becomes energetically more favorable for the excess Ni atoms to occupy \cite{Pal}. We confirm this with the results of \textit{ab~initio} calculations. However, in a quantitative comparison with experimental results theory gives a overestimation for the pinned magnetization.

To be able to reach these conclusions with experimental support, it is important to eliminate any effect arising from those other than the pinned magnetization. We find these to be effects located at the surface and sub-surface lead to additional magnetization arising from the occurrence of MnO and Ni-rich ferromagnetic Ni-Mn.

\section*{Acknowledgments}
We acknowledge funding by the German Research Foundation (DFG) within the Collaborative Research Center/Transregio (CRC/TRR) 270 (Project-No. 405553726, subprojects A04, B02 and B06). Support by the Interdisciplinary Center for Analytics on the Nanoscale of the University of Duisburg-Essen (DFG RI sources reference: RI\_00313), a DFG-funded core facility (project no. 233512597 and no. 324659309), is gratefully acknowledged. We thank Ulrich Hagemann (University of Duisburg-Essen), Ulrich Nowak (University of Konstanz) and Alfred Hucht (University of Duisburg-Essen) for helpful discussions.

\appendix
\section{First-principles calculations}
\label{Appendix}

\begin{table*}[!t]
    \caption{
    Site-resolved (in $\mu_{\rm B}$/atom) and total (in $\mu_{\rm B}$/Ni-excess-atom) spin magnetic moments~$\mu_s$ and total orbital magnetic moments~$\mu_l$ as well as MAE = $E^{110}-E^{001}$ (in meV/f.u.) of Ni$_{50+x}$Mn$_{50-x}$.
    Numbering of atoms correspond to the Fig.~\ref{fig_cell_and_Jij}a).
    }
    \label{table_magmom2}
    \centering
    \begin{ruledtabular}
    \begin{tabular}{lccccccccccc}
    \multirow{2}{*}{Ni (\si{\atpercent})} & \multicolumn{8}{c}{Site-resolved $\mu_s$ ($\mu_\text{B}$)} & \multirow{2}{*}{$\mu_s^{\rm Total}$ ($\mu_\text{B}$)} &  \multirow{2}{*}{$\mu_l^{\rm Total}$ ($\mu_\text{B}$)} & \multirow{2}{*}{MAE} \\ \cline{2-9}
    \multicolumn{1}{c}{} & Mn$_1$ & Mn$_2$ & Mn$_3$ & Mn$_4$ & Mn$_5$ & Mn$_6$ & Ni$_1$ & Ni$_2$ &  &  &  \\ \hline
    50.0625  & -3.31 & -3.31 & -3.31 & -3.31 & 3.30 & 3.30 & 0.85 & 0.01 & 4.13 & 0.13 & -0.154 \\
    53.125   & -3.21 & -3.26 & -3.30 & -3.30 & 3.27 & 3.31 & 0.64 & 0.05 & 4.83 & 0.08 &  -0.259 \\
    56.1875  & -3.12 & -3.21 & -3.28 & -3.28 & 3.24 & 3.32 & 0.50 & 0.13 & 4.91 & 0.08 &  -0.190 \\
    \end{tabular}
    \end{ruledtabular}
\end{table*}

\subsection*{Coherent potential approximation}

In~case of coherent potential approximation~(CPA), chemical disorder is modeled analytically. Calculations were performed with the help of Korringa–Kohn–Rostoker~(KKR) approach as implemented in the Munich SPR-KKR~code~\cite{Ebert-code,Ebert-2011} in the framework of atomic sphere approximation~(ASA) together with scalar relativistic corrections. The~angular momentum expansion was carried out up to $l_{max}=3$ ($f$-states). The~electronic self-consistency was assumed to be reached when the error in the potential functions dropped below~$10^{-5}$. Brillouin zone integration was carried out using the special point method with a $k$-point grid of 15~points, which corresponds to a $17 \times 17 \times 12$ mesh in the full Brillouin zone.

\begin{figure*}[!t]
    \centering
    \includegraphics[width=0.9\textwidth]{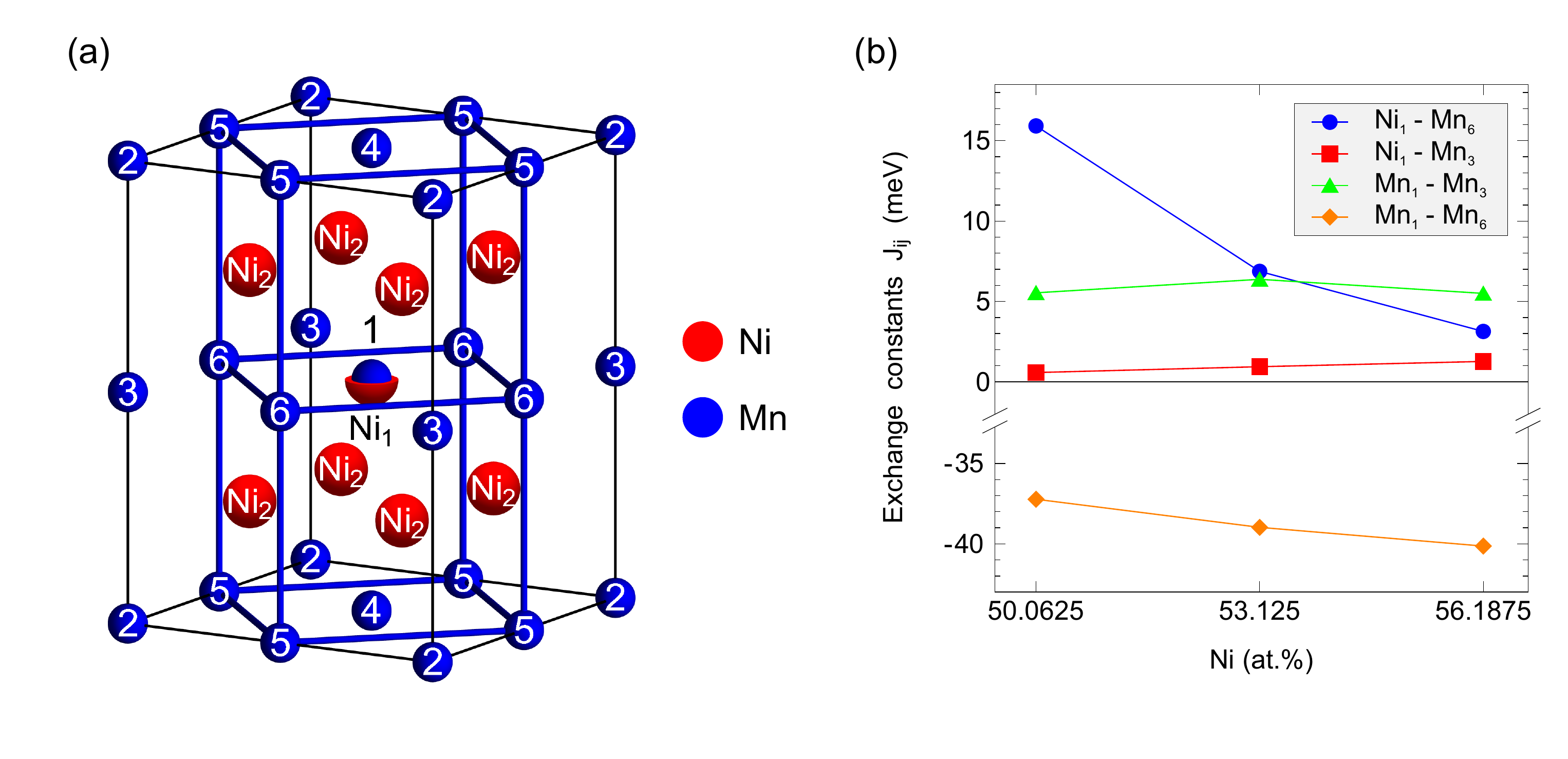}
    \caption{a)~NiMn cell used in \textit{ab~initio} calculations with CPA approach.
    Antiparallel orientation of Mn magnetic moments forms layered AFM configuration.
    b)~The~dependence of exchange constants~$J_{ij}$ of the interacting atoms~$i$ and~$i$ on Ni-excess concentration.
    $J_{ij}>0$~correspond to FM exchange, while $J_{ij}<0$ are assigned to AFM interaction.
    Numbering of atoms correspond to the computational cell depicted on the left.
    }
    \label{fig_cell_and_Jij}
\end{figure*}

To take into account systematically the effect of Ni excess on the magnetic properties, instead of equiatomic NiMn, we modeled near-equiatomic composition with slight Ni-excess on the Mn site in the middle of the cell (Mn$_1$ in Fig.~\ref{fig_cell_and_Jij}(a)). This~allows us not to exclude completely Ni$_{1}$ interactions and evaluate the influence of Ni excess on the exchange parameters.
For~the same purpose, we modeled also the intermediate composition  with equal Ni and Mn concentration in the central site (0.5, 0.5, 0.5) of the cell.
Thus, further, we will discuss the results of CPA calculations performed for Ni-concentrations of \SI{50.0625}{\atpercent} (near-equiatomic NiMn), \SI{53.125}{\atpercent}, and~\SI{56.1875}{\atpercent}.

For the above mentioned compositions, we performed calculations of the total energies, spin and orbital magnetic moments, and exchange constants.
The~exchange parameters~$J_{ij}$ define the interactions between pairs of atoms $i$ and $j$ of all different chemical types and positions as a function of the distance $r_{ij}$ between them in terms of a classical Heisenberg model Hamiltonian

\[ {\cal H}_{\rm mag}=-\sum_{i\neq j}J_{ij}\,\vec{\text{e}}_i\cdot\vec{\text{e}}_j\,, \]

\noindent where $\vec{\text{e}}_i$  and $\vec{\text{e}}_j$ describe the unit vectors of the orientation of the magnetic spin moments at sites $i$ and $j$.

Total energy calculations for the [001], [100], and [110] spin moment directions showed that [110] is slightly more favorable than the other two. Thus,~further discussion is presented for this case. Calculated spin~($\mu_s$) and orbital~($\mu_l$) magnetic moments for different Ni-excess concentrations are summarized in~Table~\ref{table_magmom2}. On~the whole, AF NiMn becomes FM with introducing Ni on the Mn site. Ni~atoms in their own sublattice~(Ni$_2$), which are nonmagnetic in the stoichiometric composition, acquire spin and orbital magnetic moments with introducing Ni-excess. Contrary to this, spin magnetic moments of extra Ni~(Ni$_1$) decrease by $\approx 1.7$~times. The~main question was the orientation of the Ni-excess magnetic moment in the Mn plane. Our~calculations show that the magnetic moment of the Ni-excess atom~(Ni$_1$) aligns parallel to the ones of nearest neighbor Mn$_6$ atoms and of Ni atoms in their planes~(Ni$_2$). Introducing Ni-excess results also in a slight decrease in $\mu_s$ of Mn$_1$ located at the center of the cell and~Mn$_2$ due to the weakening AFM Mn$_1$-Mn$_2$ and FM Mn$_1$-Mn$_1$ exchange interaction. The near-equiatomic FM Ni$_1$-Mn$_6$ exchange (blue circles in Fig.~\ref{fig_cell_and_Jij}b)) decreases by five times when Mn$_1$ is (almost) fully replaced by~Ni, which leads to a decrease in~$\mu_s^{\rm Ni_1}$.

We also calculated the magnetocrystalline anisotropy energy~(MAE) in terms of the total energy difference between two spin moment directions~\cite{Enkovaara-2002,Umetsu-2006,Gruner-2008,Edstrom-2015} as MAE~$=E_{\mathrm{tot}}^{110} - E_{\mathrm{tot}}^{001}$. The~results of MAE calculations are presented in the last column of Table~\ref{table_magmom2}. The obtained MAE~$=-0.154$~meV/f.u. of near-equiatomic NiMn agrees with the earlier theoretical study of Sakuma~\cite{sakuma1998electronic}. Introducing Ni-excess increases anisotropy of the system and the MAE becomes slightly larger in absolute value. The MAE values obtained in calculations for stoichiometric and Ni-excess NiMn are comparable with MAE for Ni$_2$MnGa~\cite{Gruner-2008,Herper-2018}.

\subsection*{Supercell approach}
Supercell calculations were performed with the help of the Vienna \textit{Ab~Initio} Simulation Package~(VASP)~\cite{Kresse-1996,Kresse-1999}. The~exchange-correlation functional was treated within the generalized gradient approximation~(GGA) following the Perdew, Burke, and Ernzerhof~(PBE)~scheme~\cite{Perdew-1996}. The~energy cut-off for the  plane wave basis set were set to~$460$~eV. The calculations converged with an energy accuracy of~$10^{-8}$~eV/atom. The~Brillouin zone integration was performed with the first order Methfessel-Paxton method using uniform Monkhorst-Pack   $4\times4\times4$ $k$-point~grid.

Modeling was performed for two structures. Firstly, we modeled 432-atom supercell of ideal NiMn by elongation of 16-atom cell with layered AFM ordering by $3 \times 3 \times 3$. Secondly, we replace one Mn in the middle of the supercell (site (0.5; 0.5; 0.5)) by Ni Ni atom. Structures used in modeling as well as the results of supercell calculations and their discussion are presented in the main text of the paper.

\end{document}